\documentclass[aps,prl,twocolumn,groupedaddress,amsmath,amssymb,showpacs,nofootinbib,floatfix]{revtex4}
\usepackage{amsfonts}
\usepackage{amsthm}
\usepackage{bm}
\usepackage{algorithmic}
\usepackage{enumerate}
\usepackage{latexsym}
\usepackage[dvips]{graphicx}

\newcommand{\beq}{\begin{equation}}
\newcommand{\eneq}{\end{equation}}
\newcommand{\beqs}{\begin{equation*}}
\newcommand{\eneqs}{\end{equation*}}
\newcommand{\bK}{{\bm K}}
\newcommand{\bKp}{{{\bm K}_+}}
\newcommand{\bKm}{{{\bm K}_-}}
\newcommand{\bb}{{\bm b}}

\begin{document}

\title{Parallel Transport of Electrons in Graphene Parallels Gravity}

\author{Andrej Mesaros$^{1}$}
\author{Darius Sadri$^{1,2}$}
\author{Jan Zaanen$^{1}$}

\affiliation{${^1}$Instituut-Lorentz for Theoretical Physics, Universiteit Leiden, P. O. Box 9506, 2300 R A Leiden, The Netherlands}
\affiliation{${^2}$Institute of Theoretical Physics, Ecole Polytechnique F\'ed\'erale de Lausanne (EPFL), CH-1015 Lausanne, Switzerland}

\begin{abstract}
Geometrically a crystal containing dislocations and disclinations can be envisaged as a `fixed frame' Cartan--Einstein space-time carrying torsion and curvature, respectively. We demonstrate that electrons in defected graphene are transported in the same way as fundamental Dirac fermions in a non-trivial 2+1 dimensional space-time, with the proviso that the graphene electrons remember the lattice constant through the valley quantum numbers. The extra `valley holonomy' corresponds to modified Euclidean symmetry generators.
\end{abstract}


\maketitle

\textit{Introduction}\,--\,The miracle of graphene is that non-relativistic electrons scattering against a lattice potential experience a low energy world in which, in non-trivial regards, they behave in ways reminiscent of Dirac's relativistic fermions: the Klein paradox, Landau quantization, the fine structure constant, and so forth. An ambitious but natural question to ponder is whether this coarse grained graphene world might mimic some aspects of gravitational structure? In Cartan's generalization of Einstein's geometrical formulation of gravity, torsion and curvature can be put in one-to-one relation with the dislocations and disclinations, the topological defects of the crystal lattice~\cite{Kleinert}. The analogy is incomplete in the sense that crystal spaces are non-diffeomorphic. The general covariance of space-time translates into the requirement that the action of the medium should be independent of arbitrary elastic deformations, something obviously violated by crystal and defected crystal geometry, which therefore correspond to spaces with a preferred metric or a `fixed frame'. In addition, in the crystal the translations and rotations of Galilean space are broken to discrete subgroups and this implies that both curvature and torsion are quantized in units of the discrete Burgers and Frank vectors, the topological invariants of the dislocations and disclinations respectively. One can nonetheless still study the transport of matter in such topologically non-trivial fixed frame backgrounds; several such analogous gravity systems have been identified, including the sound waves of superfluid $He^4$ and the nodal Fermions of the $He^3$ A phase, which percieve the hydrodynamical flow fields as geometrical (Christoffel) connection~\cite{volovikbook}.

To what extent does this analogy extend to the Dirac-like fermions in graphene? We are inspired by the previous work~\cite{Vozmediano,geomphase} demonstrating that the holonomy accumulated by electrons in graphene encircling a disclination (cone) coincides with that associated with a Dirac fermion encircling the conical singularity, the entity encapsulating curvature in 2+1 dimensional gravity. However, in order to complete the identification these earlier works added an ad-hoc $U(1)$ gauge flux to the conical singularity, acting with opposite sign on the valley quantum numbers of the graphene electrons, raising the issue of whether the identification is merely coincidental. Here we will settle these matters by focussing on the influence of dislocations, corresponding to torsion in the gravitational analogy. Torsion is a less familiar aspect of the geometrical formulation of gravity~\cite{Goeckeler,Hehl,Kibble,Kleinert}. Ignored at first by Einstein, it was introduced by Cartan~\cite{Cartan} as an a priori ingredient of a geometrical theory. It was later pointed out by Kibble~\cite{Kibble} that its inclusion becomes necessary in the presence of spinning particles, as their spin currents source torsion in a dynamical space-time, though whether torsion propagates in the space-time is a matter to be settled by observation, an as yet open question because torsional effects turn out to be too weak to be measured with present day experimental technology, but the situation is different in the `analogous' graphene system. Dislocations correspond with large 'fixed frame' localized torsion sources. We demonstrate that the holonomies associated with graphene electrons encircling dislocations resemble those coming from the most natural implementation of torsion in the connection of doubled fundamental fermions, in the case that their cones would be displaced away from zero momentum. It is just the fact that the discreteness of 'graphene geometry' is remembered exclusively by the long wavelength fermion modes by the large momenta where the Dirac cones reside, and this is surely different from the way that Planck scale discreteness (when it exists) affects fundamental fermions. We subsequently show that the mysterious $U(1)$ flux of the graphene disclination has precisely the same origin, bringing us to the conclusion that the parallel transport of electrons in graphene with dislocations and disclinations is in the long wavelength limit identical to that of Dirac fermions living at large momenta in a 2+1-d Cartan--Einstein spacetime with torsion and curvature.

\textit{Torsion in Elasticity and its Coupling to Fermions}\,--\,Within a geometric
formulation of elasticity theory, dislocations become sources of torsion~(see~[\onlinecite{LazarRev,KatanaevFirst,Kleinert}] and references therein), stemming from their translational character. Torsion $\bm{T}$ assigns a vector to an infinitesimal area element at each point in space, measuring the non-closure of a loop obtained by parallel transport of the two infinitesimals forming the ``edges'' of the given surface element along each other~\cite{Goeckeler,Hehl,Kibble,Kleinert}. The definition makes this vector completely analogous to the Burgers vector in a crystal lattice. This gravity/geometry analogy has been verified in familiar electron systems, producing results compatible with the tight-binding approach~\cite{tbtorsionSitenko,Aurell,Berrydisl}. The metric is connected to the displacement field in the crystal via (we use $\eta_{\mu\nu}$ to designate the flat Minkowski metric in a cartesian basis $\eta=\text{diag}(1,-1,-1)$, and $ds^2=g_{\mu\nu}dx^\mu dx^\nu$
for the invariant distance):
\begin{equation}
  \label{eq:1}
  g_{ij}=\eta_{ij}+\partial_iu_j+\partial_ju_i \: .
\end{equation}
Time is essentially decoupled from space in this condensed matter system ($g_{\mu 0}=0$).

Formally, torsion is defined as a vector valued 2-form on space-time,
$\bm{T}^{a}=T_{\mu\nu}^{a} dx^{\mu}\wedge dx^{\nu}$, with $\mu,\nu\in\{0,1,2\}$, and $\wedge$ the wedge product of differential forms. In this work the relevant example is of one dislocation at the origin of two dimensional space, with Burgers vector $\bb$, with a corresponding torsion
(see~\onlinecite{Kleinert,LazarRev} and references therein)
\begin{equation}
  \label{eq:5}
  \bm{T}^{a}=b^{a}\: \delta(\vec x)\: dx\wedge dy \: .
\end{equation}
The flux of this form through any area containing the origin is given by the Burgers vector $\int\!\!\int\bm{T}^{a}=b^{a}$.
As usual, the metric of Eq.~\eqref{eq:1} determines the symmetric part (in the lower indices) of the connection, i.e.~the Christoffel piece, while torsion adds additional information about parallel transport in space, being related to the asymmetric part of the connection~\cite{Kleinert,Goeckeler}, $T_{\mu\nu}^{\lambda}=\frac{1}{2}(\Gamma_{\mu\nu}^{\lambda}-\Gamma_{\mu\nu}^{\lambda})$. The geometry is consistently defined only if the Einstein-Cartan structure equations are satisfied:
\begin{subequations}
  \label{eq:2}
 \begin{align} 
  \label{eq:2a}
  \bm{R} &=d\bm{\Gamma}+\frac{1}{2}[\bm{\Gamma},\bm{\Gamma}] \: ,\\
  \label{eq:2b}
  \bm{T} &=d\beta+\bm{\Gamma}\wedge\beta \: ,
  \end{align}
\end{subequations}
where the connection $\bm{\Gamma}=\Gamma_{\mu\nu}^{\lambda}dx^\mu$ is written as a matrix valued 1-form, the curvature $\bm{R}$ is a matrix valued 2-form, and $\beta$ is a frame (i.e. $\beta^a(x)$ are the dual basis vectors of the tangent space at $x$).

Let us now define the 2+1 dimensional structure of the graphene Dirac
equation~\cite{SlonczLuttKohneffmassWall,gadget} by identifying the
Dirac matrices as $\gamma^{0}=\tau_{3}\otimes\sigma_{3}$,
$\gamma^{1}=i\tau_{3}\otimes\sigma_{2}$, and
$\gamma^{2}=-i\tau_{3}\otimes\sigma_{1}$, which satisfy the Dirac
algebra $\{\gamma^{a},\gamma^{b}\}=2\eta^{a b}$. The $\tau$ Pauli matrices act in the space of the valley index $\bK_\pm$, while the $\sigma$ Pauli matrices act on the sublattice ($A/B$) degrees of freedom.

Since spin is defined with respect to rotations acting in a tangent frame, to study the equation
of motion of a spinning particle we must introduce~\cite{Goeckeler}
an orthonormal set of basis vectors $E^a$, connected to the holonomic frame $dx^\mu$ through the vielbein (here dreibein) $dx^\mu=e_a^\mu E^a$ (and the inverse ${(e_a^\mu})^{-1}\equiv e_\mu^a$).
Intuitively since spinors are square roots of vectors, we need the square root of the metric
to define spinors on curved manifolds, and the vielbeins provide this square root
$e e \eta \sim g$; more formally they intertwine the representations of the orthogonal rotation
group and its covering spin group.
Then the relevant (zero mass) Dirac equation in a curved torsionful background takes the form
\begin{equation}
  \label{eq:3}
  i\: \gamma^a \: e_a^\mu \: \mathcal{D}_\mu \:
  \Psi=0 \: ,
\end{equation}
with the covariant derivative given via
\begin{equation}
  \label{eq:3a}
  \mathcal{D}_\mu = \left(\partial_\mu-\frac{1}{4}\omega_{\mu ab}\gamma^a\gamma^b\right) \, .
\end{equation}


The displacement field corresponding to a dislocation situated at the origin in two spatial dimensions is well known~\cite{Eshelby}, and via eq.~\eqref{eq:1} determines the metric~\cite{KatanaevWedge}
(note that $\bb$ is to be regarded as infinitesimal in the continuum theory, so that we retain only linear terms). For simplicity we take the Poisson ratio $\sigma=0$, and fix $\bb$ to point along the $x$ axis. The strategy is then to find an orthonormal basis $E^a$ on this space, i.e.
\begin{equation}
  \label{eq:4}
\eta^{ab}=e_\mu^a \: g^{\mu\nu} \: e_\nu^b \, ,
\end{equation}
and then the spin connection from eq.~\eqref{eq:2b} by using the physical input about the defect in~\eqref{eq:5}. For the basis we get
\begin{equation}
  \label{eq:6}
  \begin{pmatrix} E^1\\ E^2\end{pmatrix}=
  \begin{pmatrix}
    1-\frac{b}{2\pi r}\sin{\phi} & \frac{b}{2\pi r}\cos{\phi} \\
    \frac{b}{2\pi r}\cos{\phi} & 1-\frac{b}{2\pi r}\sin{\phi}
  \end{pmatrix}  \begin{pmatrix} dr\\ r d\phi\end{pmatrix},
\end{equation}
and $\omega^1_2=-\omega^2_1=-d\phi-\frac{b}{\pi r^2}(\cos{\phi} dr+\sin{\phi} d\phi)$. It is noteworthy that the matrix of 1-forms $\omega^\mu$ is antisymmetric as it represents the rotation of the orthogonal basis during parallel transport. The first term $d\phi$ appears simply due to the use of polar coordinates, and is responsible for a term $-\gamma^1/2r$~\cite{LCfictflux,gadget} in~\eqref{eq:3}, as well as for an unphysical delta function in the curvature obtained through~\eqref{eq:2a}, a byproduct of the singularity in the polar coordinate system.

The spin connection produces a trivial holonomy for the Dirac spinor, but a non-trivial topological action is present in the $E^a$ basis (there is in fact some freedom in the Einstein-Cartan formalism to move torsion effects between the basis one-forms and the spin-connection, obvious in eq.~\eqref{eq:2b}). The connection encodes for the integrable elastic deformation around the dislocation, and the vector field corresponding to the $\bm{\omega}$ ``potential'' in~\eqref{eq:3a} follows the deformation of the crystal due to the missing row of atoms. More quantitavely, the singularity in displacement encoding the topological defect is fully contained in $\hat{\vec{u}}_0=-\frac{b}{2\pi}\ln{(x+i y)}$, due to $\oint\hat{\vec{u}}_0=-b\hat{x}$~\cite{LandauElast}. In this case $\bm{\omega}=0$. Another example is the elastically unrelaxed displacement field in~\cite{Kleinert}, corresponding to $E^1=dx+\frac{b}{2\pi}d\phi$, $E^2=dy$, and $\bm{\omega}=0$.

The above examples are instructive in emphasizing the relation $\int\!\!\int\bm{T}^{a}=\int\!\!\int dE^a=\oint E^a=b^{a}$, where one should note that $d(d\phi)=2\pi\delta(r)\delta(\phi)dr\wedge d\phi$ due to $\phi$ not being a holonomic coordinate at the origin. Obviously, the topologically non-trivial part of $E$ will always be in the form of a $\frac{b}{2\pi}d\phi$ correction to the basis vector $E^a$ along the Burgers direction, $\bb\cdot \hat{x}^a=b$. This can easily be checked explicitly for our setup in eq.~\eqref{eq:6}, if the $E^a$ basis is rotated to $\{E^x,E^y\}$, with our previous choice of Burgers vector. The topological action of the dislocation on the Dirac electron should then be viewed as a Berry phase arising from the term
\begin{eqnarray}
  \label{eq:8}
  i\gamma^a e_a^\mu\partial_\mu\Psi&=&i\gamma^a(\delta_a^\mu+f_a^\mu)\partial_\mu\Psi \nonumber \\
  &=&i\gamma^\mu \partial_\mu\exp{\left(\int\!dx^\nu f_\nu^\mu\partial_\mu\right)}\Psi=0 \, ,
\end{eqnarray}
where $f_a^\mu(b)=f_\mu^a(-b)$ is the perturbation proportional to the Burgers vector. The non-trivial holonomy (Berry phase) is responsible for the salient feature of long range influence of the crystal defect~\cite{LCtopphase,gadget,Vozmediano}, taking the value
\begin{equation}
  \label{eq:9}
H(\bb)=e^{(\oint dx^\nu f_\nu^\mu)\partial_\mu}=e^{i\bb\cdot(-i\nabla)} \, ,
\end{equation}
where we recognize the Volterra operation of translating the wavefunction by the Burgers vector to describe the topology of a dislocation. However, the correct holonomy follows from the effect of translation by $\bb$ (which is of order of a lattice constant) on the true Bloch wavefunction~\cite{LCfictflux,gadget}, in other words
\begin{equation}
  \label{eq:10}
  H_{lattice}(\bb)=e^{i\bb\cdot\bK\tau_3} \, .
\end{equation}
The connection is striking and pleasing, because the continuum translation generator $-i\nabla$ is replaced by a translation generator $\bK\tau_3$ of the underlying lattice wavefunction, which is a finite momentum ($\bK_\pm$) state.

Eq.~\eqref{eq:9} encapsulates the essence of arguments relating the vielbein and the gauge field of Poincar\'{e} (here Euclidean) group translations in gauge theories of gravity, and one might consider a continuum-limit theory living on a background where the Euclidian group generators are modified to accommodate the lattice constant sized interactions of the defects and the finite momentum reference (Fermi) state.

\textit{Curvature and Disclinations}\,--\,In the case of disclinations, the associated curvature exists in 2+1-d as conical singularities, and has been considered in the graphene lattice~\cite{fullerene,Vozmediano,Furtado}. However, special care has to be taken to include the exchange of Fermi points, i.e.~the internal degree of freedom, that occurs for specific opening angles, by using an additional gauge field with only $\tau$ operator structure. Therefore an additional gauge field is introduced, alongside the curvature. Following the discussion in the previous section it becomes clear that it is more consistent to view the additional Fermi point effect as a change in the generator of rotations for the graphene Dirac spinor.

The correct holonomies in the presence of a disclination with the fundamental opening angles at the origin, obtained by the Volterra construction, are~\cite{fullerene,Vozmediano,LCfictflux} $H(2\pi/3)=\exp{(-i\frac{2\pi}{3}\frac{\sigma_3}{2})}$ and $H(\pi/3)=-i\tau_1\exp{(-i\frac{\pi}{3}\frac{\sigma_3}{2})}$. Note that rotating by $\pi/3$ maps the Fermi points into each other, hence the $\tau_1$ matrix. We rewrite this in an illuminating way ($\theta$ is the angle of disclination):
\begin{equation}
  \label{eq:11}
  H(\theta\equiv n\frac{\pi}{3})=e^{-i\theta(\sigma_3+3\tau_1)/2},
\end{equation}
where we see the spinor rotation (half-angle) generator $\sigma_3/2$ replaced by $(\sigma_3+3\tau_1)/2$, in order to accommodate the finite lattice constant effect due to the existence of two electron species, at finite momenta $\bK_\pm$. This is a generalization to the spinor case of the observation that the disclination holonomy is the representation of the rotation operator by the defect opening angle~\cite{ABdiscl}. It stems from the fact that the spin connection term, which produces the non-trivial holonomy in this case, is actually given by the rotation generator $\frac{1}{8}\omega_{\mu ab}[\gamma^a,\gamma^b]=\omega_{\mu 1 2}\frac{\sigma_3}{2}=\frac{\theta}{2\pi}d\phi_\mu\frac{\sigma_3}{2}$, and fixes the curvature 2--form $R^1_2=-R^2_1=d\omega=\theta\delta(\vec{x})dx\wedge dy$. Note that the $\omega_\mu$ matrix is again antisymmetric.

\textit{General Torsion Couplings}\,--\,Here we attempt to identify additional possible couplings of torsion to the specific electronic degrees of freedom in graphene, based on general considerations (see [\onlinecite{Aurell}] for a similar analysis in a different condensed matter system).

The Riemann-Cartan curved space with torsion is defined by eq.~\eqref{eq:2}, and fixed through the choice of the connection (once a tangent basis is specified), which itself provides the covariant derivative to be used in the Dirac equation, eq.~\eqref{eq:3}. This coupling of geometry to the spinor can in principle be extended by additional scalar terms containing torsion, which might follow from the choice of an action for the full gravity+matter theory~\cite{Hehl,LazarElastAct,KatanaevFirst,KatanaevWedge}, or in some cases only by an ad-hoc choice. These terms are linear in torsion at the least, and so effectively behave as a delta function potential in space (eq.~\eqref{eq:5}). Obviously this makes no contribution to a holonomy, but the discussion is interesting from a general viewpoint.

If we choose to start from a covariantized Dirac Lagrangian in 2+1 dimensions (see [\onlinecite{Hehl}] for the treatment of 3+1-d), we get an additional term in the Dirac equation $i\gamma^a\left(\nabla_a+T_{a b}^{\;\;b}\right)\Psi=0$ (written in anholonomic coordinates, with the covariant derivative becoming $\nabla_a=\partial_a-\frac{1}{4}\omega_{abc}\gamma^b\gamma^c$).  At this point we can extract all similar torsion content from the covariant derivative $\nabla_a$, by separating the antisymmetric part of the connection. Again in 2+1-d we get $\nabla_a=\tilde{\nabla}_a-\frac{1}{2}T_{a b}^{\;\;b}-\frac{1}{4}T_{a b c}\gamma^b\gamma^c$, where $\tilde{\nabla}_a$ contains only the Christoffel symbol part of the connection. The Dirac equation reads
\begin{equation}
  \label{eq:12}
i\gamma^a\left(\tilde{\nabla}_a+\frac{1}{4}T_{a b}^{\;\;b}+\frac{1}{12}\gamma_a\gamma^5_{t}\varepsilon^{bcd}T_{bcd}\right)\Psi=0 \, ,
\end{equation}
with the formally defined ``traditional'' $\gamma^5_{t}\equiv i\gamma^0\gamma^1\gamma^2=\tau_3\otimes\openone$. It seems that since the topological effect of the dislocation is present strictly in the $E^a$ basis, which stems from the singular displacement field through the metric (eqs.~\eqref{eq:1} and \eqref{eq:4}), it is enough to retain the Christoffel connection part of $\tilde{\nabla}_a$, as if there was no torsion (the additional terms in eq.~\eqref{eq:12} do not contribute). One must note, however, that torsion cannot be simply disregarded, as it is present in the space due to eq.~\eqref{eq:2b}.

Our form of $\bm{T}$ (eq.~\eqref{eq:5}) constrains the polar vector $T_{a b}^{\;\;b}=(\hat{z}\times\bb)_a\delta(\vec{x})$ to be orthogonal to the Burgers vector, and this is the only possible polar term. Considering axial vector couplings generally, in the relevant 2+1 dimensional case, there is no traditional $\gamma^{5}$ matrix which is independent of the $\gamma^{a}$ algebra, and which could be used to reduce the spinors to Weyl components, because it commutes, instead of anticommutes, with the $\gamma^{a}$. However, in the case of graphene we are dealing with a reducible representation of the Clifford algebra, built out of two irreducible ones (one at each $\bK_\pm$ Fermi point). For this case, there exists a $\gamma_{new}^{5}$ matrix, which can be defined for the present odd dimensional situation and having all the properties of $\gamma_{t}^{5}$ acting in even dimensions~\cite{Bashir}. The $\gamma_{new}^{5}$ represents the parity transformation which mixes the two irreducible representations, i.e.~in our case it must map between $\bKp$ and $\bKm$ spinor components (note that they are connected through parity, as $\bKp=-\bKm$), while in contrast the dislocation gauge coupling, which it should reproduce, acts via phase shifts without coupling the two $\bK$ points, i.e. it is of the $\tau_{3}$ form. The above observations do not prevent the appearance of terms containing $\gamma_{t}^{5}$, and the last term in eq.~\eqref{eq:12} is of such a form, but it happens to be identically zero due to the contraction $\varepsilon^{abc}T_{abc}=0$.

To further connect with the lattice dislocation coupling eq.~\eqref{eq:10}, one could consider the generalization of forming scalars making use also of the $\bK$ vector. The allowed combinations are $\varepsilon^{abc}T_{bc}^{\;\;d}K_d\gamma^5_t$ and $\varepsilon^{abc}T_{bd}^{\;\;d}K_c\gamma^5_t$, but neither is usable. The first one has the free index timelike $a=0$ (contributing a time dependent Berry phase constant in space), due to non-zero $T_{bc}^{\;\;d}$ having purely spacelike indices. The second term has the same feature ($\bK$ also has no time component), although it has the correct matrix form $\varepsilon^{abc}T_{bd}^{\;\;d}K_c\gamma^5_t=-\bb\cdot\bK\tau_3\delta(\vec{x})$.

\textit{Conclusions}\,--\,We have shown how electrons in defected graphene can be viewed as moving in a geometry with curvature and torsion, with all the topological lattice effects included in an appropriate adjustment of the underlying space symmetry generators. This is a fresh view on the subject in graphene, treating both types of defects equally, while matching them clearly with their governing symmetry sectors. We hope this perspective will aid in understanding systems with many defects, where the fact that the holonomies are non-Ableian, renders them highly non-trivial. Such a treatment must necessarily deal with intricacies of the continuum limit of the lattice description, as we have emphasized.

\textit{Acknowledgements}\,--\,This work was financially supported by the Nederlandse Organisatie voor Wetenschappelijk Onderzoek (NWO), the Stichting voor Fundamenteel Onderzoek der Materie (FOM),
and the Swiss National Science Foundation (SNSF).

\end{document}